\documentclass[10pt, conference]{IEEEtran}
\IEEEoverridecommandlockouts
\usepackage{cite}
\usepackage{amsmath,amssymb,amsfonts}
\usepackage{algorithmic}
\usepackage{graphicx}
\usepackage{textcomp}
\usepackage{xcolor}
\usepackage{hyperref}
\def\BibTeX{{\rm B\kern-.05em{\sc i\kern-.025em b}\kern-.08em
    T\kern-.1667em\lower.7ex\hbox{E}\kern-.125emX}}

\usepackage{multirow}
\usepackage{array}
    \newcolumntype{P}[1]{>{\centering\arraybackslash}p{#1}}
    \newcolumntype{M}[1]{>{\centering\arraybackslash}m{#1}}

\newcommand\T{\rule{0pt}{2.6ex}}       
\newcommand\B{\rule[-1.2ex]{0pt}{0pt}} 

\begin{document}

\title{Audio representations for deep learning in sound synthesis: A review}

\author{\IEEEauthorblockN{Anastasia Natsiou}
\IEEEauthorblockA{\textit{Technological University of Dublin} \\
Dublin, Ireland \\
anastasia.natsiou@tudublin.ie}
\and
\IEEEauthorblockN{Se\'{a}n O'Leary}
\IEEEauthorblockA{\textit{Technological University of Dublin} \\
Dublin, Ireland \\
sean.oleary@tudublin.ie}
}

\maketitle
\thispagestyle{empty}

\begin{abstract}
The rise of deep learning algorithms has led many researchers to withdraw from using classic signal processing methods for sound generation. Deep learning models have achieved expressive voice synthesis, realistic sound textures, and musical notes from virtual instruments. However, the most suitable deep learning architecture is still under investigation. The choice of architecture is tightly coupled to the audio representations. A sound's original waveform can be too dense and rich for deep learning models to deal with efficiently - and complexity increases training time and computational cost. Also, it does not represent sound in the manner in which it is perceived. Therefore, in many cases, the raw audio has been transformed into a compressed and more meaningful form using upsampling, feature-extraction, or even by adopting a higher level illustration of the waveform. Furthermore, conditional on the form chosen, additional conditioning representations, different model architectures, and numerous metrics for evaluating the reconstructed sound have been investigated. This paper provides an overview of audio representations applied to sound synthesis using deep learning. Additionally, it presents the most significant methods for developing and evaluating a sound synthesis architecture using deep learning models, always depending on the audio representation.

\end{abstract}


\section{Introduction}\label{sec:introduction}
 Sound generation algorithms synthesize a time domain waveform. This waveform should be coherent and appropriate for its intended use. These waveforms can convey complex and varied information. Deep generative networks \cite{gm_comprehensive_2020} have demonstrated great potential for such tasks having been used for the synthesis of a range of sounds, from pleasant pieces of music to natural speech \cite{huzaifah_deep_2020}. These models discover latent representations based on the distribution of the initial data and then sample from this distribution to generate new acoustic signals with the same properties as the original ones. In many cases, the deep learning models can operate along with signal processing algorithms and enhance their expression capabilities \cite{engel_ddsp_2020}\cite{valin_lpcnet_2019}.

The representation of the sound embraced by the deep neural network plays a major role on the development of the algorithm. Raw time domain audio is a rich representation which leads to massive information making the network computationally expensive and therefore slow. Compressed Time-frequency representations based on spectrograms can decrease the computer power needed but the parameter detection and synthesis of the sound is usually a challenging task and the loss of information can cause significant reconstruction error \cite{govalkar_comparison_2019}. Parameters extracted from state-of-the-art vocoders have also been proposed for deep neural network applications \cite{blaauw_neural_2017}. These parameters demonstrate a potential in marrying the deep generative models with statistical parametric synthesizers. Finally, contemporary investigations allow the network to determine the feature needed for the task \cite{engel_neural_2017}. Linguistic and acoustic features can be encoded into latent representations such as embeddings.

Apart from an overview of the audio representations existing in sound synthesis implementations, this paper additionally quotes popular schemes for conditioning a deep generative network with auxiliary data. Conditioning in generative models can control the aspects of the synthesis and lead to new samples with specific characteristics \cite{manzelli_conditioning_2018}. Furthermore, the paper highlights examples of deep generative models for audio generation applications. Deep neural networks have demonstrated remarkable progress in the field demonstrating impressive results. A final section discusses evaluation processes for synthesised sound. Subjective evaluation via listening tests are generally considered the most reliable measure of quality. However, multiple other metrics for assessing a generative model have been proposed converting both acoustic signals to intermediate representations of them to be examined. Consequently, audio representations assume an essential role not only as input data but also influence the network architecture, the conditioning technique as well as the evaluation process.

\section{Input Representations}

In the literature, numerous audio representations have been proved beneficial for audio synthesis applications. Many times, comparisons have been conducted between different forms of the sound to reveal the most appropriate representation for a specific deep learning architecture. Raw audio and time-frequency representations usually present the first attempts in such experiments. However, recent studies also look to higher level forms that offer more meaningful description such as embeddings, or multiple sound features like the fundamental frequency, loudness and features extracted by state-of-the-art vocoders such as WORLD \cite{morise_world_2016}. The table \ref{T1} summarizes the advantages and disadvantages of each sound representation.

\subsection{Waveform - Raw Audio}
The term raw audio is often used to refer to a waveform encoded using pulse code modulation (PCM). This is the sampling of a continuous waveform in time and amplitude. It represents the waveform as a sequence of numbers, each number representing an amplitude sample at a chosen sampling frequency. In order for this discrete sequence of samples to capture all the necessary information, the highest frequency in the signal should adhere to the Nyquist-Shannon theorem \cite{shannont_communication_1949}. According to this theorem, only frequencies of less than half the sampling frequency can be reproduced from the sampled signal. A typical sampling frequency for audio applications is 44.1kHz. Each real number is assigned to the approximate fixed value in a finite set of discrete numbers. The most common levels for quantization are stored in 8 bits (256 levels), 16 bits (65536 levels) and 24 bits (16.8 million levels). Therefore, a sound with a duration of one second sampled at 44.1kHz generates 44100 samples. This representations is considered extremely informative even for deep learning networks.

In order for the outcome of the deep learning model to be more effective, a pre-processing step can be used to reduce the quantization range of the raw audio. Many research approaches \cite{oord_wavenet_2016}\cite{oord_parallel_2017}\cite{binkowski_high_2019} apply $\mu$-law to decrease the possible values of each prediction. $\mu$-law is presented in Eq. \ref{mulaw} where $-1<x<1$ and $\mu$ equals the number of levels created after the transformation.

\begin{equation}\label{mulaw}
f(x)=sgn(x)\frac{ln(1+\mu|x|)}{ln(1+\mu)}
\end{equation}

Although non-linear quantization processes such as $\mu$-law received much attention the last years, the majority of the existing papers use a normalized high resolution signal as input \cite{kim_flowavenet_2019}. Finally, other applications include linear quantization of the input waveform \cite{mehri_samplernn_2017}\cite{yamamoto_probability_2019} and different designs for most and less significant bits \cite{kalchbrenner_efficient_2018}.

\subsection{Spectrograms}
A spectrogram is a time/frequency visual representation of sound. A spectrogram can be obtained via the Short Time Fourier Transform (STFT), where the Fourier Transform is applied to overlapping segments of the waveform. The Discrete Fourier Transform (DFT) is presented by the equation Eq. \ref{dft} for $k=0,1,..,N-1$ where N is the number of samples and k is the number of segments. The spectrogram uses just the absolute values of the STFT, discarding the phase. This type of spectrograms has been used in many  by a variety of papers \cite{wang_tacotron_2017}\cite{neekhara_expediting_2019}\cite{arik_fast_2019}.

\begin{equation}\label{dft}
X(k)=\sum_{n=0}^{N-1}x(n)e^{-j\omega_{k}n}
\end{equation}

Apart from the original spectrogram, deep learning architectures have also  experimented with non linear spectrograms such as mel-spectrograms \cite{prenger_waveglow_2018}\cite{peng_non-autoregressive_2020}\cite{aouameur_neural_2019}\cite{ren_fastspeech_2019}\cite{ren_fastspeech_2021}\cite{liu_wavetts_2020}\cite{vasquez_melnet_2019} or Constant-Q Transformations (CQT) \cite{huang_timbretron_2019}. The mel-spectrogram is generated by the application of perceptual filters on the DFT called mel-filter bands. The most common formula for encoding mel-filter bands is presented by Eq. \ref{eq_mel} where $f$ is the frequency in Hertz. However, other models have captured the perceptual transformation applying a linear transformation until 1kHz and a logarithmic above this threshold.

\begin{equation}\label{eq_mel}
mel = 2595log_{10}(1+\frac{f}{700})
\end{equation}

CQT is another time-frequency representation where the frequencies are geometrically spaced. The centre frequencies of the filters  are calculated from the result of the formula $\omega_k = 2^{\frac{k}{b}}\omega_0$ where $k = 1, 2, .. k_{max}$ and b is a constant number. The bandwidth of each frequency then comes as $\delta_k = \omega_{k+1} - \omega_k = \omega_k(2\frac{1}{b}-1)$ and therefore the frequency resolution is determined by the Eq. \ref{cqt} where Q is the quality factor. 

\begin{equation}\label{cqt}
Q = \frac{\omega_k}{\delta_k} = (2^{\frac{1}{b}}-1)^{-1}
\end{equation}

CQT is a representation with different frequency resolution in low and high frequencies. However, the phase part is discarded and the representation is in most of the cases irreversible. Following this argument, Velasco et al \cite{velasco_constructing_2011} proposed an invertible CQT based on nonstationary Gabor frames. Another variation of CQT is rainbowgrams. Rainbowgrams proposed by Engel et al \cite{engel_neural_2017} using colors to encode time derivatives of the phase.

In addition, more complicated spectrogram-based representations have also been investigated. GANSynth \cite{engel_gansynth_2019} conducted experiments with numerous spectrograms including scaled logarithmic amplitude and phase of the STFT, increased resolution of the original spectrogram or applied mel-filters. Also, they examined an Instantaneous Frequency based spectrogram where the phase of the STFT is scaled and unwraped (add $-\pi$) and then the finite difference between the frames is computed. Other applications \cite{ping_clarinet_nodate}\cite{donahue_adversarial_2019}, also, made comparisons between raw audio and spectrogram to uncover the most functional representation for their deep learning model.

\subsection{Acoustic Features}
Overcoming the wealth of acoustic information presenting in a sound waveform, various studies extract perceptual features from the original signal. These acoustic features can be represented by phoneme inputs \cite{wang_style_nodate}, fundamental frequency and spectral features \cite{wang_neural_2019} or multiple information such as the velocity, instrument, pitch and time \cite{defossez_sing_nodate}. Other implementations have included cepstral coefficients \cite{valin_lpcnet_2019}\cite{subramani_vapar_2020} or a variety of linguistic and acoustic features \cite{juvela_glotnetraw_2019}\cite{binkowski_high_2019}. Finally, widely recommended parameters have also been extracted by the WORLD vocoder \cite{blaauw_neural_2017}\cite{yang_statistical_2017}\cite{henter_deep_2018}.

\subsection{Embeddings}
Embeddings initially introduced by Natural Language Processing (NLP) in order to convert a word or sentence into a real-valued vector. This approach assisted the process of text in deep learning applications by inserting the property of closer embeddings in vector space to encode words with similar meaning. The same approach has been adopted by sound processing to reduce the dimensionality of the signal \cite{bitton_neural_2020}\cite{peng_non-autoregressive_2020}, enhance the timbre synthesis \cite{engel_ddsp_2020} or even generate a more interpretable representation \cite{esling_generative_2018}\cite{esling_universal_2019} to effectively extract parameters for a synthesizer. In \cite{engel_neural_2017} an autoencoder generates a latent representation to condition a WaveNet model while Dhariwal et al \cite{dhariwal_jukebox_2020} implemented three separate encoders to generate vectors with different temporal resolutions.

\subsection{Symbolic}
In music processing, the term symbolic refers to the use of representations such as Musical Instrument Digital Interface (MIDI) or piano rolls. MIDI is a technical standard that describes a protocol, a digital interface or the link for the simultaneous operation between multiple electronic musical instruments. A MIDI file demonstrates the notes being played in every time step. Usually this file consists of information of the instrument being played, the pitch and its velocity. MidiNet \cite{yang_midinet_2017} is one of the most popular implementations using MIDI to generate music pieces.

Piano roll constitutes a more dense representation of MIDI. A piece of music can be represented by a binary $N \times T$ matrix where N is the number of playable notes and T is the number of timesteps. In MuseGAN \cite{dong_musegan_2017}, Generative Adversarial Networks (GANs) have been applied for music generation using multiple-track piano-roll representation. Also, in DeepJ \cite{mao_deepj_2018}, they scaled the representation matrix between 0 and 1 to capture the note's dynamics. The most notable disadvantage of symbolic representations is that holding a note and replaying a note are demonstrated by the same representation. To differentiate these two stages, DeepJ included a second matrix called \textit{replay} along with the original matrix \textit{play}.

\begin{table*}
    \centering
    \caption{Overview of sound representations}
    \label{T1}
      \begin{tabular}{||P{2cm} |P{1.6cm}|P{5cm}|P{5cm}|P{2cm}||}
         \hline
         Representation & Papers & Advantages & Disadvantages & Comments \T\B\\ 
         \hline\hline
         
        \multirow{3}{*}{Waveform}  
        & \T \cite{oord_wavenet_2016}\cite{oord_parallel_2017}\cite{binkowski_high_2019}\cite{kim_flowavenet_2019}\cite{mehri_samplernn_2017} \cite{yamamoto_probability_2019}\cite{kalchbrenner_efficient_2018} \B
         & 
         \multirow{3}{5cm}{-Completely describes the waveform. \\ -Directly generates the output waveform.}
        & 
        \multirow{3}{5cm}{-Computationally expensive. \\ -Unstructured representation that does not reflect sound perception.}
        &
        \multirow{3}{2cm}{\centering Used as input or conditional representation}\\ [1ex]
         \hline

         \multirow{4}{*}{Spectrograms}  & \T \cite{wang_tacotron_2017}\cite{neekhara_expediting_2019}\cite{arik_fast_2019}\cite{prenger_waveglow_2018}\cite{peng_non-autoregressive_2020}\cite{aouameur_neural_2019}\cite{ren_fastspeech_2019}\cite{ren_fastspeech_2021}\cite{liu_wavetts_2020}\cite{vasquez_melnet_2019}\cite{huang_timbretron_2019}\B
         & 
         \multirow{4}{5cm}{-Interpretable representations that are related to sound perception. \\ -Easy to illustrate/plot.}
        & 
        \multirow{4}{5cm}{-Typically phase is discarded meaning it is not directly invertible.}
        &
        \multirow{4}{2cm}{\centering Used as input or conditional representation}\\ [1ex]
         \hline
         \multirow{4}{*}{Acoustic features} & \T \cite{wang_style_nodate}\cite{wang_neural_2019}\cite{defossez_sing_nodate}\cite{valin_lpcnet_2019}\cite{subramani_vapar_2020}\cite{juvela_glotnetraw_2019}\cite{binkowski_high_2019}\cite{blaauw_neural_2017}\cite{yang_statistical_2017}\cite{henter_deep_2018} \B
         & 
         \multirow{4}{5cm}{-Compressed, descriptive representation of aspects of sound.}
        & 
         \multirow{4}{5cm}{-Difficult to synthesize waveforms with long term coherence. \\ -Typically don’t fully specify sounds.}
        & 
         \multirow{4}{2cm}{\centering Mostly used for \\ conditioning}\\ [1ex]
         \hline
         
         \multirow{3}{2cm}{\centering Latent\\ representations} 
         &  \T \cite{engel_ddsp_2020}\cite{peng_non-autoregressive_2020}\cite{bitton_neural_2020}\cite{esling_generative_2018}\cite{esling_universal_2019}\cite{engel_neural_2017}\cite{dhariwal_jukebox_2020} \B
         & 
         \multirow{3}{5cm}{-Similar sounds lead to smaller distance in multi-dimensional space. \\ -Compressed representation. }
        &
         \multirow{3}{5cm}{-Losses in decoding. \\ -Can be difficult to interpret. }
            & 
            \multirow{3}{*}{Mostly VAEs}\\ [1ex]
         
         \hline
         \multirow{2}{*}{Symbolic} & \T \cite{yang_midinet_2017}\cite{dong_musegan_2017}\cite{mao_deepj_2018} \B
        & 
         \multirow{2}{5cm}{-Meaningful description of musical content. }
        &
         \multirow{2}{5cm}{-Very high level description of audio.}
            & \\ [1ex]
         \hline
        \end{tabular}
  \end{table*}

\section{Conditioning Representations}\label{conditioning_representations}

Neural networks are able to generate sound based on the statistical distribution of the training data. The more uniform the input data to the network is, the more natural outcome can be achieved. However, in cases where the amount of training data is not sufficient, additional data with similar properties can be included by applying conditioning methods. Following these techniques, the generated sound can be conditioned on specific traits such as speaker's voice \cite{zhao_wasserstein_2018}\cite{vasquez_melnet_2019}, independent pitch \cite{engel_ddsp_2020}\cite{jin_fftnet_2018}\cite{subramani_vapar_2020}, linguistic features \cite{arik_deep_2017}\cite{kalchbrenner_efficient_2018} or latent representations \cite{valin_lpcnet_2019}\cite{dong_musegan_2017}. Instead of one-hot-embeddings, some implementations have also used a confusion matrix to capture a variation of emotions \cite{henter_deep_2018}, while others provided supplementary positional information of each segment conditioning music to the artist or genre \cite{dhariwal_jukebox_2020}. After training, the user is able to decide between the conditioning properties of the synthesised sound.

\subsection{Additional Input}
The simplest strategy for applying conditioning to deep learning architectures is by including auxiliary input data while training. Two types of conditioning have been proposed, global and local \cite{oord_wavenet_2016}\cite{kong_diffwave_2021}. In global conditioning, additional latent representations can be appended across all training data. Global conditioning can encode speaker's voice or linguistics features. Local conditioning usually refers to supplementary timeseries with lower sampling rate than the original waveform or even mel-spectrograms, logarithmic fundamental frequency or auxiliary pitch information.

WaveNet has achieved one of the most effective strategies for conditioning deep neural networks \cite{boilard_literature_nodate}. Therefore, later sound generation schemes adopted a WaveNet network for conditioning. The majority of these works conditioned their model to spectrograms \cite{ping_clarinet_nodate}\cite{shen_natural_2018}\cite{arik_deep_2017}\cite{ping_deep_2018}\cite{wang_style_nodate}\cite{li_neural_2019}\cite{huang_timbretron_2019}\cite{angrick_speech_2019} while others included linguistic features and pitch information \cite{oord_parallel_2017}\cite{arik_deep_nodate}, phoneme encodings \cite{blaauw_neural_2017}, features extracted from the STRAIGHT vocoder \cite{tamamori_speaker-dependent_2017} or even MIDI representations \cite{hawthorne_enabling_2019}.

Although it has been proven that convolutional networks are capable of effective conditioning, other architectures can use auxiliary input data as well. Recurrent neural networks such as LSTMs have been adopted conditioning as frame-level auxiliary feature vectors \cite{ling_waveform_2018} or as one-hot representation encoding music style \cite{mao_deepj_2018}. Autoencoders can be conditioned including additional input to the encoder \cite{aouameur_neural_2019}\cite{subramani_vapar_2020} but also as input only to the decoder \cite{bitton_neural_2020}.

\subsection{Input to the Generator}
Generative Adversarial Networks (GANs) consist of two separate networks, the Generator and the Discriminator. Following the fundamental properties of GANs, the Generator converts random noise to structured data while the Discriminator endeavors to classify a signal as original or generated. For applying conditioning in GANs, the most common technique constitutes a biased input to the Generator. In sound synthesis, a well established conditioning method includes the mel-spectrogram as input to the Generator \cite{yamamoto_parallel_2020}\cite{yamamoto_probability_2019}\cite{neekhara_expediting_2019}. This way, the synthesised sound is not just a product of a specific distribution but it also obtains desirable properties. For example, it can be enforced to conditioning on predetermined instrument or voice. Furthermore, a Generator conditioned on spectrograms can also be used as a vocoder \cite{kumar_melgan_2019}. In addition to the mel-spectrogram, other implementations have been conditioned on raw audio \cite{pascual_segan_2017}, one-hot vectors to encode musical pitch \cite{engel_gansynth_2019}, linguistic features \cite{binkowski_high_2019}, or latent representations to identify speaker \cite{donahue_end--end_2021}.

\subsection{Other}
At last, other variations of conditioning have been introduced as well. Kim et al \cite{kim_flowavenet_2019} adjusted conditioning through the loss function. They estimated an auxiliary probability density using mel-spectrograms for local conditioning. Pink et al \cite{ping_waveflow_nodate} applied bias terms in every layer of the convolutional network using also mel-spectrograms. Extra bias to the network has been also proposed by \cite{rao_grapheme--phoneme_2015} to encode linguistic features while in \cite{engel_neural_2017} every layer was biased with a different linear projection of the temporal embeddings. In \cite{yang_statistical_2017} linguistic features were added to the output of each hidden layer in the Generator while in \cite{yang_midinet_2017} a new network was introduced by the name conditionerCNN to work along with the Generator encoding chords for melody generation. Finally, Juvela et al \cite{juvela_glotnetraw_2019} conducted a comparative study of conditioning methods.

\section{Methods}
During recent years, deep learning models have significantly contributed to research on sound generation. Using a variety of deep learning algorithms, multiple representations have been applied. The most common architectures include autoregressive methods, variational autoencoders (VAE), adversarial networks and normalising flows. However, many approaches can fall in more than one category.

\subsection{Autoregressive}
Autoregressive models define a category of generative models where every new sample in a sequence of data depends on previous samples.  Autoregressive deep neural networks can be represented by architectures that demonstrate this continuation implicitely or explicitely. Conventional methods that implicitely indicate a time-related manner are the recurrent neural networks. These models are able to recall previous data dynamically using complex hidden state. SampleRNN \cite{mehri_samplernn_2017} is one well established research work that applies hierarchical recurrent neural networks such as GRU and LSTM on different temporal resolutions for sound synthesis. In order to illustrate the temporal behaviour of the network, Mehri et al conducted experiments to test the model's memory by injecting one second of silence between two random sequential samples. Other significant papers on autoregressive models using recurrent neural networks are WaveRNN \cite{kalchbrenner_efficient_2018}, MelNet \cite{vasquez_melnet_2019} or LPCNet \cite{valin_lpcnet_2019}. In WaveRNN, they introduced a method for reducing the sampling time by using a batch of short sequences instead of a unique long sequence while maintaining high quality sound.

Generative models where the synthesis of the sequential samples follows a conditional probability distribution like the one in Eq.\ref{chain} are able to explicitly demonstrate temporal dependencies. 

\begin{equation}\label{chain}
p(X) = \prod_{t-1}^{T}p(x_{t}|x_{1},...,x_{t-1})
\end{equation}
WaveNet \cite{oord_wavenet_2016} presents the most influential architecture of explicit autoregressive models. 
The probability distribution can be imitated by a stack of convolutional layers. However, to improve efficiency, the sequential data passes through a stack of dilated causal convolutional layers where the input data are masked to skip some dependencies. Following a similar scheme, FFTNet \cite{jin_fftnet_2018} takes advantage of convolutional networks mimicking the FFT algorithm while upsampling the input data. However, to clear up the confusion, convolutional networks do not always lead to autoregressive models \cite{arik_fast_2019}.

Autoregressive models have been initially proposed for sequential data. Therefore, in sound synthesis, raw audio is ordinarily used as the input representation. However, many auxiliary representations have been applied conditioning the audio generation on a variety of properties. More details about conditioning techniques for autoregressive models have been already presented in Section \ref{conditioning_representations}. 

Since autoregressive models can be applied on sequential data, they are well established in sound generation related topics. Autoregressive models are easy to train and they can manipulate data in real time. Furthermore, convolutional-based models can be trained in parallel. Nevertheless, although these models can be paralleled during training, the generation is sequential and therefore slow. Synthesised data are affected only by previous samples, providing half way dependencies. Finally, the generation can be consistent to specific properties for a definite number of samples and the outcome often lacks global structure.

\subsection{Normalizing Flow}
Normalizing flows constitute a family of generative models consisting of multiple simple distributions for transforming input data to latent representations. A sequence of simple, invertible and computationally inexpensive mappings $z \backsim p(z)$  can model a reversible complex one. This complex transformation is presented in Eq. \ref{flow1} and the inverse can be achieved by repeatedly changing the variables as shown in Eq. \ref{flow2}. The mapping functions, then, can be parametrised by a deep neural network.

\begin{equation}\label{flow1}
x = f_0 \circ f_1 \circ ... \circ f_k(z)
\end{equation}
\begin{equation}\label{flow2}
z = f_{k}^{-1} \circ f_{k-1}^{-1} \circ ... \circ f_{0}^{-1}(x)
\end{equation}

WaveGlow \cite{prenger_waveglow_2018}, a flow-based generative network, can synthesise sound from its mel-spectrogram. By applying an Affine Coupling Layer and a 1x1 Invertible Convolution, the model aims to maximise the likelihood of the training data. The implementation has been proposed by NVIDIA and it is able to generate sound in real time. Insightful alternatives have also been proposed on normalising flows by using only a single loss function, without any auxiliary loss terms \cite{kim_flowavenet_2019} or by applying dilated 2-D convolutional layers \cite{ping_waveflow_nodate}. 

Finally, in order to reduce the number of repeated iterations needed by normalising flows, they have been merged with autoregressive methods. This architecture manages to increase the performance of autoregressive models since the sampling can be processed in parallel. Using Inverse Autoregressive Flows (IAF), Oord et al increased the efficiency of WaveNet \cite{oord_parallel_2017}. Their implementation follows a "probability density distillation" where a pre-trained WaveNet model is used as a teacher and scores the samples a WaveNet student outputs. This way, the student can be trained in accordance with the distribution of the teacher. A similar approach has been adopted by ClariNet \cite{ping_clarinet_nodate}, where a Gaussian inverse autoregressive flow is applied on WaveNet to train a text-to-wave neural architecture.

\subsection{Adversarial Learning}
Unlike the Inverse Autoregressive Flows where a pre-trained teacher network assist a student model, in adversarial learning, two neural networks match against each other in a two-player minimax game. The fundamental architecture of Generative Adversarial Networks (GANs) is based on two models, the Generator (G) and the Discriminator (D). The Generator maps a latent representation to the data space. In a vanilla GAN, the Generator maps random noise to a desirable representation. For sound synthesis this representation could be raw audio or spectrogram. This desired representation, original or generated, is used as input to the Discriminator which is trained to distinguish between real and fake data. The maximum benefit from GANs is acquired when the Generator produces perfect data and the Discriminator is not able to differentiate between real and fake data.

From a more technical point of view, the Discriminator is trained using only the distribution of the original data. Its purpose is to maximise the probability of identifying real and generated data. On the other hand, the Generator is trained through the Discriminator. Information about the original distribution of the dataset are concealed from it and its aim is to minimise the error of the Discriminator. This minimax game can be summarised by the Eq. \ref{gans}.

\begin{equation}\label{gans}
\begin{aligned}
\min_{G}\max_{D}V(D,G) = \mathbb{E}_{x \backsim p_{data}(x)}[\log D(x)] \\
+ \mathbb{E}_{z \backsim p_{z}(z)}[\log(1-D(G(z)))]
\end{aligned}
\end{equation}

On the field of sound generation a variety of implementations have been proposed using numerous representations. In \cite{engel_gansynth_2019}, spectrograms were generated using upsampling convolutions for fast generation while in \cite{donahue_adversarial_2019}, they investigated whether waveform or spectrograms are more effective for GANs applying the Wasserstein loss function. In Parallel WaveGAN \cite{yamamoto_parallel_2020}, a teacher-student scheme was adopted using non-autoregressive WaveNet in order to improve WaveGAN's efficiency. Yamamoto et al \cite{yamamoto_probability_2019}  applied GANs using a IAF generator optimised by a probability density distillation algorithm. Also, in GAN-TTS \cite{binkowski_high_2019}, they examined an ensemble of Discriminators to generate acoustic features using the Hinge loss function along with \cite{kumar_melgan_2019}\cite{donahue_end--end_2021}. Lastly, GANs have also been applied in a variety of  applications such as text-to-speech applications \cite{neekhara_expediting_2019}, speech synthesis \cite{oyamada_generative_2018}\cite{saito_statistical_2018}, speech enhancement \cite{pascual_segan_2017} or symbolic music generation \cite{dong_musegan_2017}.

\subsection{Variational Autoencoders}

An autoencoder is one of the fundamental deep learning architectures consisting of two separate networks, an encoder and a decoder. The encoder compresses the input data into a latent representation while the decoder synthesises data from the learned latent space. The original scheme of an autoencoder was initially created for dimensionality reduction purposes. Although theoretically the decoder bear some resemblance to the generator of GANs, the model is not well qualified for the synthesis of new examples. The network endeavors to reconstruct the original input, therefore it lacks of expressiveness.

To use autoencoders as generative models, variational autoencoders have been proposed \cite{kingma_auto-encoding_2014}. In this architecture, the encoder first models a latent distribution and then the network samples from the distribution to generate latent examples. The success of the variational autoencoders is mostly based on the Kullback–Leibler (KL) divergence used as a loss function. The encoder introduces a new distribution $q(z|X)$ to estimate $p(z|X)$ as much as possible by minimising the KL divergence. The complete loss function is demonstrated in the Eq. \ref{vae} where the first term (called \textit{reconstruction loss}) is applied on the final layer and the second term (called \textit{regularization loss}) adjusts the latent layer.

\begin{equation}\label{vae}
\mathcal{L} = \mathbb{E}_{z \backsim q(z|X)} [\log p(X|z)] - D_{KL}[q(z|X)||p(z)]
\end{equation}

Many variations of VAE have been applied for sound generation topics. In \cite{engel_ddsp_2020} they used VAE with feedforward networks and an additive synthesiser to reproduce monophonic musical notes. In \cite{pandey_new_2019} and \cite{bitton_neural_2020} they applied convolutional layers while in \cite{subramani_vapar_2020} a Variational Parametric Synthesiser was proposed using a conditional VAE. 

A modification of variational autoencoders proposed for music synthesis is VQ-VAE \cite{dhariwal_jukebox_2020}. In this approach, the network is trained to encode the input data into a sequence of discrete tokens. Jukebox introduces this method to flatten the data and process it using autoregressive Transformers. 

\section{Evaluation}
Although in the last decade generative models presented significant improvements, a definitive evaluation process still remains an open question. Many mathematical metrics have been proposed for perceptually evaluating the generative sound and usually a transformation to another audio representation have been adopted. However, despite the numerous attempts, none of these metrics are as reliable as the subjective evalution of human listeners.

\subsection{Perceptual Evaluation}
Human evaluation usually accounts for the mean opinion score between a group of listeners. To conduct the study, many researchers used crowdMOS \cite{ribeiro_crowdmos_nodate}, a user-friendly toolkit for performing listening evaluations. As well as the mean opinion score, a confidence interval is also been computed. Furthermore, in order to attract an accountable number of subjects with specific characteristics, Amazon Mechanical Turk has been widely used. In many cases, raters have been asked to pass a hearing test \cite{prenger_waveglow_2018}, keep headphones on \cite{kumar_melgan_2019}\cite{oord_wavenet_2016}, or only native speakers for evaluating speech have been asked \cite{yamamoto_parallel_2020}\cite{kumar_melgan_2019}\cite{li_neural_2019}.

In these mean opinion score tests, subjects have been asked to rate a sound in a five-point Likert scale in terms of pleasantness \cite{prenger_waveglow_2018}, naturalness \cite{oord_wavenet_2016}\cite{binkowski_high_2019}\cite{donahue_end--end_2021}, sound quality \cite{kalchbrenner_efficient_2018} or speaker diversity \cite{donahue_adversarial_2019}. In addition, subjects have been requested to express a preference between sounds of two generative models hearing the same pitch \cite{engel_gansynth_2019} or speech \cite{kalchbrenner_efficient_2018}\cite{oord_wavenet_2016}. Finally, for evaluating WaveGAN \cite{donahue_adversarial_2019}, humans listened to digits between one to ten and were asked to indicate which number they heard.

\subsection{Number of Statistically-Different Bins}
The Number of Statistically-Different Bins (NDB) is a metric for unconditional generative models in order to estimate the diversity of the synthesised examples. Clustering techniques are applied on the training data creating cells of similar properties. Then, the same algorithm tries to categorise the generated data into the cells. If a generated example does not belong to a predefined cluster, then the generated sound is statistically significantly different. 

GANSynth \cite{engel_gansynth_2019} used $k$-means to map the log spectrogram of the generated sound into $k = 50$ Voronoi cells. As well as Mean Opinion Score and the Number of Statistically-Different Bins, GANSynth also used Inception Score, Pitch Accuracy and Pitch Entropy and Frechet Inception Distance for evaluation purposes. The rest of the metrics will be analysed in the following sections. A similar set of evaluation metrics has also been adopted by \cite{kong_diffwave_2021} including NDB.

\subsection{Inception Score}
The Inception Score (IS) is another perceptual metric which correlates with human evaluation and is mostly adopted by GANs. For the Inception Score, a pre-trained Inception classifier is applied to the output of the generative model. In order to measure the diversity of the synthesised data, the IS calculates the KL divergence between the model scores $P(y|x)$ and the marginal distribution $P(y)$ as it can be expressed by the Eq.\ref{IS} for every possible class \cite{salimans_improved_2016}. The IS is maximised when the generated examples belong to only one class and every class is predicted equally often.

\begin{equation}\label{IS}
IS = exp(\mathbb{E}_{x}D_{KL}(P(y|x)||P(y)))
\end{equation}

In \cite{engel_gansynth_2019} and \cite{engel_neural_2017}, a pitch classifier is trained on spectrograms of the NSynth dataset while in WaveGAN \cite{donahue_adversarial_2019}, the classifier is trained on normalised log mel-spectrograms having zero mean and unit variance. Finally, metrics like Pitch Accuracy and Pitch Entropy or a nearest neighbour technique have been adopted by GANSynth and WaveGAN respectively to further evaluate the efficiency of their Inception Score. Finally, in \cite{kong_diffwave_2021} they also applied a modified inception score. 

\subsection{Distances-based measurements}
This evaluation category includes metrics that measure the distance between representations of the original data and the distribution of the generated examples. Binkowski et al. proposed two distance-based metrics, the Fr\'{e}chet DeepSpeech Distance (FDSD) and the Kernel DeepSpeech Distance (KDSD) \cite{binkowski_high_2019} for evaluating their text-to-speech model. The two metrics make use of the the Fr\'{e}chet distance and the Maximum Mean Discrepancy respectively on audio features extracted by a speech recognition model.

The Fr\'{e}chet or 2-Wasserstein distance has been proposed by other research papers as well. Engel et al \cite{engel_gansynth_2019} applied the Fr\'{e}chet Inception Distance on features extracted by a pitch classifier while Kilgour et al \cite{kilgour_frechet_2019} used this distance to measure the intensity of a distortion in generated sound. However, although many researchers report successful results using 2-Wasserstein, Donahue et al \cite{donahue_end--end_2021} reported that a similar evaluation metric did not produce a desirable outcome in their experiments.

Distances-based measurements have also been investigated individually by separate parameter estimations. In \cite{engel_ddsp_2020}  distances between the generated loudness and fundamental frequency of synthesised and training data are used.

\subsection{Spectral Convergence}
The Spectral Convergence expresses the mean difference between the original and the generated spectrogram. It has been applied by \cite{dhariwal_jukebox_2020}\cite{arik_fast_2019}\cite{tamamori_speaker-dependent_2017}\cite{bitton_neural_2020} in order to evaluate their synthesised music. The Spectral Convergence can be expressed by the Eq.\ref{spectral_convergence} which is also identified as the minimization process of the Griffin-Lim algorithm.

\begin{equation}\label{spectral_convergence}
SC = \sqrt{\frac{\sum_{n,m}|S(n,m)-\widetilde{S}(n,m)|^2}{\sum_{n,m}S(n,m)}}
\end{equation}

\subsection{Log Likelihood}
A final evaluation metric includes a Negative Log Likelihood (NLL) \cite{kalchbrenner_efficient_2018}\cite{mehri_samplernn_2017} and an objective Conditional Log Likelihood (CLL) \cite{kim_flowavenet_2019} usually measured in bits per audio sample.

\section{Conclusion}
The choice of audio representation is one of the most significant factors in the development of deep learning models for sound synthesis. Numerous representations have been proposed by previous researchers focusing on different properties. Raw audio is a direct representation demanding notable memory and computational cost. It is also not considered for evaluating purposes since different waveforms can perceptually produce the same sound. Spectrograms can overcome some of the disadvantages of raw audio and have been considered as an alternative for training as well as for evaluation. However, reconstructing the original sound from its spectrogram is a challenging task since it may produce sound suffering from distortions and lack of phase coherence. Recently, other audio representations have received much attention such as latent representations, embeddings and acoustic features but they all require a powerful decoder. The choice of audio representation is still very much dependent on the application.

\section{Acknowledgments}

This publication has emanated from research supported in part by a grant from Science Foundation Ireland under Grant number 18/CRT/6183. For the purpose of Open Access, the author has applied a CC BY public copyright licence to any Author Accepted Manuscript version arising from this submission’.


\bibliographystyle{ieeetr}
\bibliography{LitReview}
\end{document}